\documentclass[conference, 10pt, final]{IEEEtran}
\usepackage[nolist]{acronym} 
\usepackage{tikz}
\usepackage{pgfplots}
\usepgfplotslibrary{groupplots}
\usepackage{graphicx}
\usepackage{subfloat}
\usepackage{caption}
\usepackage{subcaption}
\usepackage{float}
\pgfplotsset{compat=newest}
\usepackage{units}
\usepackage{amsmath, amsbsy, amssymb}
\usepackage{tikzscale}
\usetikzlibrary{arrows}
\usepackage{color}
\usepackage{epigraph}
\usepackage{circuitikz}
\usepackage{multirow}
\usepackage{booktabs}
\usepackage[plainruled]{algorithm2e}
\usepackage[group-separator={,}]{siunitx}
\definecolor{darkgreen}{rgb}{0.125,0.5,0.169}
\usetikzlibrary{shapes,arrows}
\usepackage{marvosym}
\usepackage{bbm}
\usepackage{mathtools}

\usepgfplotslibrary{fillbetween}
\usetikzlibrary{matrix}
\usetikzlibrary{dsp,chains}
\usetikzlibrary{shapes.geometric}

\tikzset{>=latex}

\usetikzlibrary{external}
\renewcommand{\vec}[1]{\mathbf{#1}}

\newcommand{\cv}{\vec{c}}

\newcommand{\Hm}{\vec{H}}

\newcommand{\Ym}{\vec{Y}}

\begin{acronym}
 \acro{CSI}{channel state information}
 \acro{UE}{user equipment}
 \acro{UL}{uplink}
 \acro{BS}{basestation}
 \acro{TDD}{time division duplex}
 \acro{FDD}{frequency division duplex}
 \acro{ECC}{error-correcting code}
 \acro{MLD}{maximum likelihood decoding}
 \acro{HDD}{hard decision decoding}
 \acro{IF}{intermediate frequency}
 \acro{RF}{radio frequency}
 \acro{SDD}{soft decision decoding}
 \acro{NND}{neural network decoding}
 \acro{CNN}{convolutional neural network}
 \acro{ML}{maximum likelihood}
 \acro{GPU}{graphical processing unit}
 \acro{BP}{belief propagation}
 \acro{LTE}{Long Term Evolution}
 \acro{BER}{bit error rate}
 \acro{SNR}{signal-to-noise-ratio}
 \acro{ReLU}{rectified linear unit}
 \acro{BPSK}{binary phase shift keying}
 \acro{QPSK}{quadrature phase shift keying}
 \acro{AWGN}{additive white Gaussian noise}
 \acro{MSE}{mean squared error}
 \acro{LLR}{log-likelihood ratio}
 \acro{MAP}{maximum a posteriori}
 \acro{NVE}{normalized validation error}
 \acro{BCE}{binary cross-entropy}
 \acro{CE}{cross-entropy}
 \acro{BLER}{block error rate}
 \acro{SQR}{signal-to-quantisation-noise-ratio}
 \acro{MIMO}{multiple-input multiple-output}
 \acro{OFDM}{orthogonal frequency division multiplex}
 \acro{RF}{radio frequency}
 \acro{LOS}{line of sight}
 \acro{NLoS}{non-line of sight}
 \acro{NMSE}{normalized mean squared error}
 \acro{CFO}{carrier frequency offset}
 \acro{SFO}{sampling frequency offset}
 \acro{IPS}{indoor positioning system}
 \acro{TRIPS}{time-reversal IPS}
 \acro{RSSI}{received signal strength indicator}
 \acro{MIMO}{multiple-input multiple-output}
 \acro{ENoB}{effective number of bits}
 \acro{AGC}{automated gain control}
 \acro{ADC}{analog to digital converter}
 \acro{ADCs}{analog to digital converters}
 \acro{FB}{front bandpass}
 \acro{FPGA}{field programmable gate array}
 \acro{JSDM}{Joint Spatial Division and Multiplexing}
 \acro{NN}{neural network}
 \acro{IF}{intermediate frequency}
 \acro{LoS}{line-of-sight}
 \acro{NLoS}{non-line-of-sight}
 \acro{DSP}{digital signal processing}
 \acro{AFE}{analog front end}
 \acro{SQNR}{signal-to-quantisation-noise-ratio}
 \acro{SINR}{signal-to-interference-noise-ratio}
 \acro{ENoB}{effective number of bits}
 \acro{AGC}{automated gain control}
 \acro{PCB}{printed circuit board}
 \acro{EVM}{error vector mangnitude}
 \acro{CDF}{cumulative distribution function}
 \acro{MRC}{maximum ratio combining}
 \acro{MRP}{maximum ratio precoding}
 \acro{MRT}{maximum ratio transmission}
 \acro{DeepL}{deep-learning}
 \acro{DL}{deep learning}
 \acro{SISO}{single-input single-output}
 \acro{SGD}{stochastic gradient descent}
 \acro{CP}{cyclic prefix}
 \acro{MISO}{Multiple Input Single Output}
 \acro{LMMSE}{linear minimum mean square error}
 \acro{ZF}{zero forcing}
 \acro{USRP}{universal software radio peripheral}
 \acro{RNN}{recurrent neural network}
 \acro{GRU}{gated recurrent unit}
 \acro{LSTM}{long short-term memory}
 \acro{NTM}{neural turing machine}
 \acro{DNC}{differentiable neural computer}
 \acro{TCN}{temporal convolutional network}
 \acro{FCL}{fully connected layer}
 \acro{MANN}{memory augmented neural network}
 \acro{RNN}{recurrent neural network}
 \acro{DNN}{dense neural network}
 \acro{FIR}{finite impulse response}
 \acro{BPTT}{back-propagation through time}
 \acro{GAN}{generative adversarial network}
 \acro{ELU}{exponential linear unit}
 \acro{tanh}{hyperbolic tangent}
 \acro{BICM}{bit-interleaved coded modulation}
 \acro{OTA}{over-the-air}
 \acro{IM}{intensity modulation}
 \acro{DD}{direct detection}
 \acro{RL}{reinforcement learning}
 \acro{SDR}{software-defined radio}
 \acro{WGAN}{Wasserstein generative adversarial network}
 \acro{BMD}{bit-metric decoding}
 \acro{BMI}{bit-wise mutual information}
 \acro{LDPC}{low-density parity-check}
 \acro{IDD}{iterative demapping and decoding}
 \acro{IEDD}{iterative estimation, demapping and decoding}
 \acro{JSD}{Jensen-Shannon divergence}
 \acro{MMSE}{minimum mean square error}
 \acro{FFT}{fast Fourier transform}
 \acro{IFFT}{inverse fast Fourier transform}
 \acro{QAM}{quadrature amplitude modulation}
 \acro{EMD}{earth mover's distance}
 \acro{TDL}{tapped delay line}
 \acro{KL}{Kullback-Leibler}
  \acro{IDFT}{inverse discrete Fourier transform}
 \acro{DFT}{discrete Fourier transform}
 \acro{PDP}{power delay profile}
 \acro{ISI}{inter-symbol interference}
 \acro{LS}{least squares}
 \acro{V2X}{Vehicle-to-Everything}
 \acro{TDDL}{time-distributed dense layer}
 \acro{APP}{a posterior probability}
 \acro{WSSUS}{wide-sense stationary uncorrelated scattering}
 \acro{EXIT}{extrinsic information transfer}
 \acro{AE}{autoencoder}
 \acro{FEC}{forward error correction}

\end{acronym}

\definecolor{mittelblau}{RGB}{0, 126, 198}
\definecolor{violettblau}{cmyk}{0.9, 0.6, 0, 0}
\definecolor{rot}{RGB}{238, 28 35}
\definecolor{apfelgruen}{RGB}{140, 198, 62}
\definecolor{gelb}{RGB}{1, 221, 0}
\definecolor{orange}{RGB}{244, 111, 33}
\definecolor{pink}{RGB}{237, 0, 140}
\definecolor{lila}{RGB}{128, 10, 145}
\definecolor{hellgrau}{RGB}{224, 224, 224}
\definecolor{mittelgrau}{RGB}{128, 128, 128}
\definecolor{dunkelgrau}{RGB}{80,80,80}
\definecolor{anthrazit}{RGB}{19, 31, 31}

\definecolor{medpurple}{RGB}{147,112,219}
\definecolor{light_brown}{RGB}{160,82,45}
\definecolor{uni_gelb}{cmyk}{0, 0.1, 1, 0}
\pgfplotscreateplotcyclelist{corporate colours markers}{%
rot, every mark/.append style={fill=.!80!rot},mark=*\\%
mittelblau, every mark/.append style={fill=.!80!mittelblau},mark=square*\\%
apfelgruen, every mark/.append style={fill=.!80!apfelgruen},mark=triangle*\\%
orange, mark=star\\%
pink, every mark/.append style={fill=.!80!pink},mark=diamond*\\%
violettblau, every mark/.append style={fill=.!80!violettblau},mark=otimes*\\%
lila, mark=|\\%
gelb, every mark/.append style={fill=.!80!gelb},mark=pentagon*\\%
hellgrau, mark=text,text mark=p\\%
anthrazit, mark=text,text mark=a\\%
}

\pgfplotscreateplotcyclelist{corporate colours markers double}{%
rot, every mark/.append style={fill=.!80!rot},mark=*\\%
rot, dashed, every mark/.append style={fill=.!80!rot,solid},mark=*\\%
mittelblau, every mark/.append style={fill=.!80!mittelblau},mark=square*\\%
mittelblau, dashed, every mark/.append style={fill=.!80!mittelblau,solid},mark=square*\\%
apfelgruen, every mark/.append style={fill=.!80!apfelgruen},mark=triangle*\\%
apfelgruen, dashed, every mark/.append style={fill=.!80!apfelgruen,solid},mark=triangle*\\%
orange, mark=star\\%
orange, dashed, mark=star\\%
pink, every mark/.append style={fill=.!80!pink},mark=diamond*\\%
pink, dashed, every mark/.append style={fill=.!80!pink,solid},mark=diamond*\\%
violettblau, every mark/.append style={fill=.!80!violettblau},mark=otimes*\\%
violettblau, dashed, every mark/.append style={fill=.!80!violettblau,solid},mark=otimes*\\%
lila, mark=|\\%
lila, dashed, mark=|\\%
gelb, every mark/.append style={fill=.!80!gelb},mark=pentagon*\\%
gelb, dashed, every mark/.append style={fill=.!80!gelb,solid},mark=pentagon*\\%
}

\IEEEoverridecommandlockouts

\begin{document}

\title{Adaptive Neural Network-based OFDM Receivers}

\author{\IEEEauthorblockN{Moritz Benedikt Fischer$^{1}$, Sebastian D\"orner$^{1}$, Sebastian Cammerer$^{2}$,\\Takayuki Shimizu$^{3}$, Hongsheng Lu$^{3}$, and Stephan ten Brink$^{1}$\\}

\IEEEauthorblockA{
$^{1}$ Institute of Telecommunications, University of Stuttgart, Pfaffenwaldring 47, 70569 Stuttgart, Germany \\
\{fischer,doerner,tenbrink\}@inue.uni-stuttgart.de\\
$^{2}$ NVIDIA, Fasanenstra{\ss}e 81, 10623 Berlin, Germany, scammerer@nvidia.com\\
$^{3}$ InfoTech Lab, Toyota Motor North America, \{takayuki.shimizu,hongsheng.lu\}@toyota.com\\
}

\thanks{This work has been supported by Toyota Motor North America and by the Federal Ministry of Education and Research of the Federal Republic of Germany through the FunKI project under grant 16KIS1187.}
}

\maketitle

\begin{abstract}
We propose and examine the idea of continuously adapting state-of-the-art \ac{NN}-based \ac{OFDM} receivers to current channel conditions.
This online adaptation via retraining is mainly motivated by two reasons:
First, receiver design typically focuses on the universal optimal performance for a wide range of possible channel realizations.
However, in actual applications and within short time intervals, only a subset of these channel parameters is likely to occur, as \emph{macro} parameters, e.g., the maximum channel delay, can assumed to be static.
Second, \emph{in-the-field} alterations like temporal interferences or other conditions out of the originally intended specifications can occur on a practical (real-world) transmission.
While conventional (filter-based) systems would require reconfiguration or additional signal processing to cope with these unforeseen conditions, \ac{NN}-based receivers can learn to mitigate previously \emph{unseen} effects even after their deployment.
For this, we showcase on-the-fly adaption to current channel conditions and temporal alterations solely based on recovered labels from an outer \ac{FEC} code without any additional piloting overhead.
To underline the flexibility of the proposed adaptive training, we showcase substantial gains for scenarios with static channel macro parameters, for out-of-specification usage and for interference compensation.
\end{abstract}

\section{Introduction}
The ongoing trend of applying \acp{NN} to signal processing tasks for communication systems has led to the demonstration of substantial improvements when compared to conventional systems for a wide range of applications \cite{honkala2020deeprx,samuel2017deep,li2018power}.
Especially when focusing on recent results of \ac{NN}-based \ac{OFDM} receivers \cite{honkala2020deeprx, aoudia2020end, Fischer_2021}, where implementations showed comparable, or sometimes even better performance than conventional state-of-the-art baselines%
, there is reason to believe that \ac{NN}-based components will play a significant role in future beyond 5G systems \cite{Toward6G_Hoydis_2021}.
Based on the assumption that trainable components will be present in future receivers, we want to discuss the opportunity of online retraining during operation to further adapt to current channel conditions.

Conventionally, receiver algorithms are designed offline, where they are optimized for best performance on comprehensive channel models, focusing on universal optimal performance.
At the same time, these channel models are optimized to mimic the expected average behavior of the real-world channel as accurately as possible.
This also holds for \ac{NN}-based receivers, which are typically trained offline on a data-set representing an ensemble of channel realizations generated by the same underlying channel model.
Training \ac{NN}-based receivers could also be done using measured data, but this entails several difficulties as the measurements must cover a wide range of different channel conditions to enable the NN to generalize to the task, and are therefore expensive.
Thus, initially training \ac{NN}-based receivers on generated data is advantageous for generalization due to the randomness introduced by stochastic channel models.
This has been done in \cite{aoudia2020end, Fischer_2021} and results in similar or even superior performance compared to conventional \ac{LMMSE}-based systems, when also evaluated on the same stochastic channel models.

\begin{figure}[t]
\begin{center}
\input{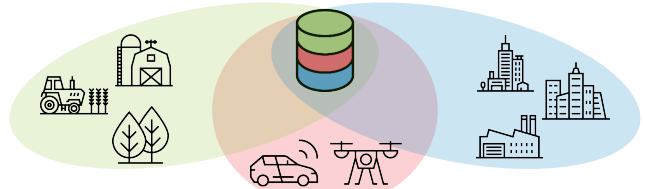}
\end{center}
\vspace{-2mm}
\caption{Visualization of sub-ensembles representing various channel conditions within a universal training data-set.}
\label{fig:channel_ensemble}
\vspace{-4mm}
\end{figure}
However, in an actual real-world system and within a short period of time, only a subset of these universal channel conditions occurs.
The receiver rather observes sub-ensembles of conditions, sketched schematically in Fig.~\ref{fig:channel_ensemble}, depending on the area of current operation (rural, urban, city) or situation (velocity, interference).
As these \emph{macro} conditions only change slowly, compared to signal processing from the receiver's point of view, we want to investigate the impact of retraining the initially universally optimized receiver for the actual channel conditions.
From a deep learning perspective, this approach can be seen as a deliberate overfitting, since we propose to retrain the receiver with only the latest data available.

In the following, we show by using the example of \ac{NN}-based \ac{OFDM} receivers, that re-optimizing to the current channel conditions leads to gains compared to the universally optimized system in corner cases and demonstrate that retrained receivers can also adapt to initially unseen channel conditions and channel alterations like interference.
The paper is structured as follows: Sec.~\ref{sec:system_setup} introduces the channel model and \ac{OFDM} system.
In Sec.~\ref{sec:RNN} details on the applied \ac{RNN}-based \ac{OFDM} receiver and the adaptive retraining process are given.
Finally, Sec.~\ref{sec:results} presents simulation results and Sec.~\ref{sec:conclusion} concludes the main findings.

\section{System Setup}
\label{sec:system_setup}

The ideal channel data to showcase the advantages of online retraining would be temporally continuous ``in-the-field'' measurements of \ac{CSI} for \ac{UE} trajectories covering various different channel conditions.
An equally potent alternative to measured data could be ray-tracing-based \ac{CSI}, simulated for \ac{UE} trajectories within large spatially consistent areas.
Unfortunately, to the best of our knowledge, neither of both data sources satisfying these requirements are currently available.
This is why we rely on a modified Jakes' and Clarke's oriented time-varying and frequency-selective stochastic channel model for our simulations.
By sensitively manipulating the stochastic model's parameters, e.g., maximum channel delay, \ac{PDP} or \ac{UE} velocity, we can generate stochastic sub-ensembles of channel realizations representing the different channel conditions as simplistically visualized in Fig.~\ref{fig:channel_ensemble}.

\subsection{Channel Model and OFDM System}
We consider a tapped-delay line channel model with time-varying channel impulse response $h\left(t, \tau\right)$.
The time-varying channel impulse response is defined as
\begin{equation}
    h\left(t, \tau\right) = \sum_{\ell=0}^{L-1} a_{\ell}\left(t\right)\delta\left(\tau - \tau_{\ell}\right)
\end{equation}
where $L$ is the number of resolvable multipath-components, i.e., taps, $a_{\ell}$ is the complex time-varying gain of the ${\ell}$th tap, $\tau_{\ell}$ is the delay of the ${\ell}$th tap\footnote{In the following it is assumed that the delay of the first tap is \unit[0]{ns} and that the delay time is equally spaced with $\nicefrac{1}{B}=\unit[100]{ns}$.} and $\delta\left(.\right)$ is the Dirac delta function.
For each channel realization, these multipath-components $a_{\ell}$ are randomly generated to hold a certain average power $p_{\ell} = \operatorname{E}\left[|a_{\ell}|^2\right]$ while their absolute value $|a_{\ell}|$ is Rayleigh distributed. %
This average power $p_{\ell}$ of the ${\ell}$th multipath-compenent is assumed to follow an exponentially decaying \ac{PDP}.
Each channel tap is therefore weighted during its generation with the weight $b_{\ell} = \sqrt{p_{\ell}}$ computed by 
\begin{equation}
    \label{eq:exp_dec}
    b_{\ell} = \frac{1}{\gamma}\sqrt{1-\beta}\cdot \beta^{\nicefrac{{\ell}}{2}} \in \mathbb{R}, \qquad {\ell} = 0,1,...,L-1
\end{equation}
where the factor $\gamma$ is chosen such that $\sum_{\ell}|b_{\ell}|^2=1$ and ${0<\beta<1}$ is a variable decay parameter.
The Fourier transform of the channel impulse response $h\left(t, \tau\right)$ then yields the channel transfer function $H \left( t,f \right)$.

We assume that the considered \ac{OFDM} transmission system operates on frames of $n_\mathrm{T}$ consecutive \ac{OFDM} symbols with parameters given in Tab.~\ref{Tab:Scenario}.
Each \ac{OFDM} symbol consists of $N_{\mathrm{Sub}}$ symbols -- either data-carrying or pilot-carrying -- that are transmitted in parallel over the $N_\mathrm{Sub}$ subcarriers.
The transmitted information bits $\mathbf{u}$ %
are encoded and interleaved into the sequence $\mathbf{c}$ of length $n_{\mathrm{d}}\cdot m$ using an 5G NR compliant \ac{LDPC} code  \cite{5G_Code_2018} of length $n=1296$ bit. Here, $n_\mathrm{d}$ denotes the number of transmitted data-carrying symbols within a frame and each data symbol carries the information of $m$ bits (e.g., $m=4$ for a 16 \ac{QAM}).
For the simulation in frequency domain it is assumed that a sufficiently long \ac{CP} is applied and \ac{ISI} is not present. %
Let $\mathbf{X} \in \mathbb{C}^{n_{\mathrm{T}}\times N_{\mathrm{Sub}}}$ be the transmitted symbols.
After the removal of the \ac{CP} the received symbols $\mathbf{Y}\in\mathbb{C}^{n_{\mathrm{T}}\times N_{\mathrm{Sub}}}$ are given by
\begin{equation}
    \label{eq:received_symbols}
    \mathbf{Y} = \mathbf{H} \circ \mathbf{X} + \mathbf{N}
\end{equation}
where $\circ$ denotes the element-wise multiplication, $\mathbf{H}\in \mathbb{C}^{n_{\mathrm{T}}\times N_{\mathrm{Sub}}}$ is the channel matrix and $\mathbf{N}\in \mathbb{C}^{n_{\mathrm{T}}\times N_{\mathrm{Sub}}}$ is the \ac{AWGN} matrix. 
By sampling $H\left(t,f\right)$ according to the \ac{OFDM} system parameters given in Tab.~\ref{Tab:Scenario} we end up with the channel matrix $\mathbf{H}$ of the current frame.
The elements $N_{k,n}$ of the noise matrix $\mathbf{N}$ are independent and identically complex Gaussian distributed according to 
$N_{k,n}\sim \mathcal{CN}\left(0, \sigma^2\right)$ where
$\sigma^2$ denotes the noise power per element.
The task at receiver side is to equalize and demap the received symbols $\mathbf{Y}$. %
Finally, the obtained soft bit estimates are decoded by a \ac{BP} decoder. %

\subsection{Iterative LMMSE Baseline}
As a state-of-the-art baseline system, we employ a receiver based on the \ac{IEDD} principle. %
It consists of a data-aided \ac{LMMSE} channel estimator, a (soft-decision) \ac{APP} demapper and a \ac{BP} decoder that iterates and exchanges soft bit information with the estimator and the demapper.
For further details the interested reader is referred to \cite{aoudia2020end} and the references therein. 

\section{Adaptive RNN-based OFDM Receiver}
\label{sec:RNN}

To demonstrate the advantages of adaptive retraining we consider a trainable \ac{RNN}-based \ac{OFDM} receiver. %
Similar to \cite{honkala2020deeprx,aoudia2020end}, it combines the tasks of channel estimation, equalization and soft-demapping within a single \ac{NN}.%

\subsection{Neural Network Structure and Training}

\begin{figure}[t]
\begin{center}
\input{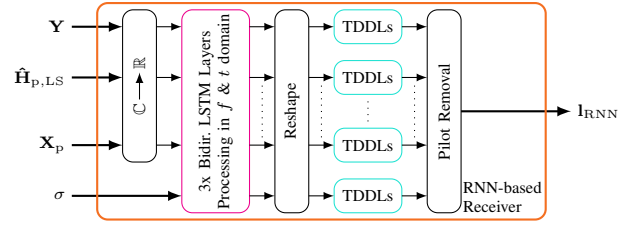}
\end{center}
\vspace{-2mm}
\caption{Block diagram of the \ac{RNN}-based \ac{OFDM} receiver.}
\label{fig:RNN_structures}
\vspace{-5mm}
\end{figure}

Fig.~\ref{fig:RNN_structures} provides an overview of the applied \ac{NN} model which is based on the structure that has been used in \cite{Fischer_2021} for the task of channel estimation. %
The RNN maps the received symbols $\mathbf{Y}$ to a soft bit estimation, interpreted as \acp{LLR} $\mathbf{l}_{\mathrm{RNN}}\in \mathbb{R}^{n_{\mathrm{d}}\cdot m}$. %
Besides $\mathbf{Y}$, it also takes the transmitted pilot symbols $\mathbf{X}_\mathrm{p} \in \mathbb{C}^{ n_\mathrm{T} \times N_{\mathrm{Sub}}}$%
, the \ac{LS} channel estimates $\hat{\mathbf{H}}_\mathrm{p,LS}\in \mathbb{C}^{n_{\mathrm{T}} \times N_{\mathrm{Sub}}}$ at pilot positions and the noise standard deviation $\sigma$ into account. %
The complex-valued inputs are split into their real and imaginary parts and the noise standard deviation is broadcasted for the whole frame to match the input tensor shape, so that all inputs can be stacked to one large input tensor.
Similar to \cite{Fischer_2021}, the core element of the \ac{RNN} cell are three bidirectional \ac{LSTM} layers that primarily process the input.
The first \ac{LSTM} layer operates along the input's frequency dimension.
Next, the output's frequency and time dimension are permuted causing the second \ac{LSTM} layer to operate in time dimension.
Finally, the time dimension and the frequency dimension of the second layer's output are again permuted so that the third \ac{LSTM} layer again processes along the frequency dimension of the frame.  
Subsequently, %
the \ac{RNN} cell's output is reshaped and processed by two \acp{TDDL}. %
Here, every element of the two-dimensional resource grid of the frame is processed separately by these \acp{TDDL} using shared weights. %
The \ac{LSTM} cells are applied with TensorFlow's default settings using \ac{tanh} activations, the first \ac{TDDL} uses \acp{ReLU} and the second \ac{TDDL} has no activation function. %
In this work, we use 64 units within each \ac{LSTM} layer, the first \ac{TDDL} consists of 8 neurons and the second \ac{TDDL} uses $m$ neurons, i.e., the RNN outputs $m$ values for every position in the resource grid. %
After removing the output values at pilot positions, the \ac{RNN}'s reshaped output $\mathbf{l}_{\mathrm{RNN}} \in \mathbb{R}^{n_\mathrm{d}\cdot m}$ can be de-interleaved and utilized by the outer \ac{BP} decoder. %

Training of the described \ac{RNN} is carried out in a supervised manner utilizing \ac{SGD} and \ac{BPTT}.
During training (initial as well as re-training) the Adam optimizer \cite{Kingma2014} with a learning rate of $\eta = 0.001$ is used to minimize the \ac{BCE} loss between estimations $\mathbf{l}_{\mathrm{RNN}}$ and labels $\cv$.
The \ac{RNN}-based receiver is initially trained with universal randomly generated channel realizations from the stochastic channel model for a vast range of different channel parameters.
This kind of initial training results in an universal and robust generalization and allows the \ac{RNN}-based receiver to implicitly gather knowledge of the channel only through data-driven training \cite{Fischer_2021}.
The exact parameters used for initial training are summarized in Tab.~\ref{Tab:Training_Parameters}. 

	\begin{table}[t]
		\centering
		\vspace{0.03in}
		\caption{Parameters for Initial (Universal) Training}
		\vspace{-1mm}
		\begin{tabular}{l|l}
			\toprule
			Parameter & Value  \\
			\midrule
            Epochs / It. per epoch / BS & 100 / 1000 / 128 \\
			Velocity $v$& $\unitfrac[0]{km}{h}- \unitfrac[200]{km}{h}$ \\
			Signal-to-noise-ratio (SNR) & $\unit[8]{dB} - \unit[30]{dB}$\\
			Number of channel taps $L$ & Ep. 1-50: 4-10; Ep. 51-100: 1-14\\
			\ac{PDP} & Exp. decaying with $10\operatorname{log_{10}}\left(\frac{p_{L-1}}{p_0}\right)$\\& $=\unit[-13]{dB}$ and equally spaced\\%Exp. decaying with the power \\&in the last resolvable path being\\ & $\unit[13]{dB}$ lower than the power of\\& the first path and equally spaced\\  %
			
			\bottomrule	
		\end{tabular}
		\label{Tab:Training_Parameters}
	\vspace{-5.5mm}
	\end{table}

\subsection{Adaptive Retraining via On-the-fly Label Recovery}
\label{sec:retraining}

In order to allow the \ac{RNN}-based \ac{OFDM} receiver to adapt to current channel conditions, it has to be retrained periodically.
To enable a single retraining step, a data-set consisting of multiple recorded OFDM frames (holding inputs $\mathbf{Y}$, $\mathbf{X}_\mathrm{p}$, $\hat{\mathbf{H}}_\mathrm{p,LS}$ and $\sigma$) and the corresponding labels, being the originally transmitted interleaved coded bits $\mathbf{c}$, must be collected.
As the labels $\mathbf{c}$ are required for supervised training, they must either be retrieved by the transmission of pilot-based training sequences (and are thereby known at the receiver side) or via on-the-fly label recovery, as presented in \cite{schibisch2018online}.
Whereas pilot-based training sequences would cause a rate loss, the approach proposed in \cite{schibisch2018online} recovers the labels on-the-fly via the outer \ac{FEC} after the decoder has corrected the received bits.
Thus, there is no additional rate loss and these labels usually come for free as most systems rely on \acp{FEC}.

\begin{figure}[t]
	 	\begin{center}
	 		\input{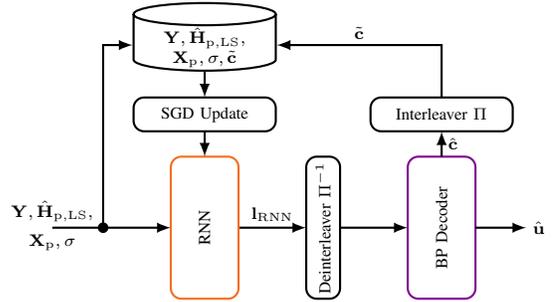}
	 	\end{center}
	 	\vspace{-2mm}
	 	\caption{Block diagram of the retraining process for NN-based receiver adaptation via on-the-fly label recovery \cite{schibisch2018online}.}
	 	\label{fig:on_the_fly_label_recovery}
	 	\vspace{-8mm}
	 \end{figure}

To demonstrate the feasibility of on-the-fly label recovery for the task of RNN retraining, we only use labels recovered by the LDPC code after 20 iterations of BP decoding.
The block diagram in Fig.~\ref{fig:on_the_fly_label_recovery} depicts the individual processing steps that allow retraining with recovered labels. %
Therefore, the \ac{RNN} processes the received symbols as described above and outputs an \ac{LLR} for each transmitted bit. 
These \acp{LLR} $\mathbf{l}_{\mathrm{RNN}}$ are then de-interleaved and further processed by the \ac{BP} decoder. %
In normal operation, the decoder makes a final decision on the received information bits $\hat{\mathbf{u}}$ after several iterations of \ac{BP} decoding.
But, in order to build up a labeled data-set for retraining, at the same time the decoder also outputs its information on the coded bits $\hat{\mathbf{c}}$, i.e., a hard decision on the final variable nodes.
These coded bits $\hat{\mathbf{c}}$ are then interleaved to $\tilde{\mathbf{c}}$ and stored together with the corresponding inputs.

If enough tuples of inputs and labels are recovered to form a sufficiently large retraining data-set, an update step using supervised \ac{SGD} is performed, aiming to reduce the \ac{BCE} loss.
However, one drawback of the described label recovery approach is, that even after sufficient decoding, not all labels can be recovered correctly by a \ac{FEC} code.
This is why we consider a codeword's error syndrome in combination with the current \ac{SNR} to define a threshold for labels that are stored in the retraining data-set, while samples above the threshold are discarded.
Similar to the findings in \cite{schibisch2018online} we saw improved performance after retraining even with partly erroneous labels.
If the number of erroneous labels exceeded a certain level we saw a degradation after retraining.
But, this can be avoided by defining the threshold conservatively.%

\section{Simulation Results}
\label{sec:results}

\begin{table}[t]
\centering
\vspace{0.03in}
\caption{OFDM and Channel Model Parameters}
\vspace{-1mm}
\begin{tabular}{l|l}
	\toprule
	Parameter & Value  \\
	\midrule
	Number of subcarriers $N_{\mathrm{Sub}}$ & 64 \\
	Frame length $n_{\mathrm{T}}$& 36 \\
	Carrier frequency $f_{\mathrm{c}}$ & $\unit[5.9]{GHz}$  \\
	Symbol duration including \ac{CP} $T_{\mathrm{S}}$ & $\unit[8]{\mu s}$  \\
	Length of the \ac{CP} &$\unit[1.6]{\mu s}$ \\
	Bandwidth $B$ & $\unit[10]{MHz}$\\
	Data symbol constellation & 16 QAM, $m=4$ bit per symbol \\
	Pilot structure/arrangement & Rectangular/Grid \\
	Pilot symbol distance & $d_{\mathrm{T}}=15$, $d_\mathrm{F}=5$\\
	\ac{PDP} & Exp. decaying with\\ &$10\operatorname{log_{10}}\left(\frac{p_{L-1}}{p_0}\right)=\unit[-13]{dB}$\\ %
	LDPC code & $R_{\mathrm{C}} = \nicefrac{1}{2}$, $n = \unit[1296]{bit}$\\ %
	\bottomrule	
\end{tabular}
\label{Tab:Scenario}
\vspace{-4mm}
\end{table}

To evaluate the effects of adaptive retraining we simulate the performance of various receiver setups in three different scenarios.
For each scenario we assume certain channel conditions, simulated by channel model parameters, to be static for a short period of time.
Within this time period, which shall represent the \emph{current} channel, we gather retraining data via on-the-fly label recovery as described in Sec.~\ref{sec:retraining}, perform a retraining step of the RNN-based receiver and then evaluate the performance on the same channel conditions.
For the following simulation results, a retraining step was executed after 32 batches with 50 frames of input-label-tuples per batch were collected.
With the general simulation parameters given in Tab.~\ref{Tab:Scenario}, this translates to a label recovery time period of $\unit[0.4608]{s}$ and, thereby, sets a lower bound (neglecting time for retraining computations) for periodic retraining steps to track channel alterations.
To limit the amount of erroneous labels within a recovered retraining data-set, we empirically defined the threshold according to the codeword's error syndrome in a way that at least $82\%$ of the parity-checks of the recovered labels have to be fulfilled by a batch to be used for retraining.
In addition, a batch is only used for retraining if the \acs{SNR} $\nicefrac{E_{\mathrm{b}}}{N_0}$ is larger than $\unit[7]{dB}$, resulting in basically no retraining in the low SNR regime.\footnote{Pilot sequence-based labels are required for retraining in the low SNR regime, as recovered labels based on FEC suffer from high error rates.}
Also, each recovered batch is only used once for an SGD weight update iteration and one retraining step is performed separately for every evaluation point at different SNR.
For each scenario the performance is measured by the \ac{BER} after forward error correction (post-FEC) and the following receiver systems are analyzed: 

\begin{itemize}
    \item \textbf{Universal \ac{RNN}}: Non-iterative RNN-based receiver, %
    initially trained with the universal parameters summarized in Tab.~\ref{Tab:Training_Parameters}, complemented by 20 iterations of BP decoding.
    
    \item \textbf{Adapted \ac{RNN}:} Non-iterative \ac{RNN}-based receiver, initially trained with the universal parameters in Tab.~\ref{Tab:Training_Parameters}, that is adapted to the current channel via one retraining step using on-the-fly recovered labels. Also complemented by 20 iterations of BP decoding.
    
    \item \textbf{\ac{LMMSE} \ac{IEDD}}: Conventional \ac{LMMSE} \ac{IEDD} baseline system %
    utilizing an autocorrelation matrix that is matched to the channel (genie knowledge of channel model parameters). %
    The \ac{BP} decoder executes 5 iterations %
    before feedback is provided to estimator and demapper. %
    In total $4\times 5=20$ iterations of BP decoding are executed.

    \item \textbf{Perfect Knowledge IDD}: Lower limit of the achievable \ac{BER} assuming perfect knowledge of the channel and utilizing an iterative receiver, i.e., exploiting \ac{IDD}. 
    Here, feedback is provided to the demapper after every iteration of \ac{BP} decoding and $\Hm$ is known. %
    In total  $20\times 1 = 20$ iterations of BP decoding are executed.
    \vspace{-2mm}
\end{itemize}

\subsection{Corner Case (Sub-Ensemble) Scenario}

\begin{figure}[t]
    \begin{center}
        \input{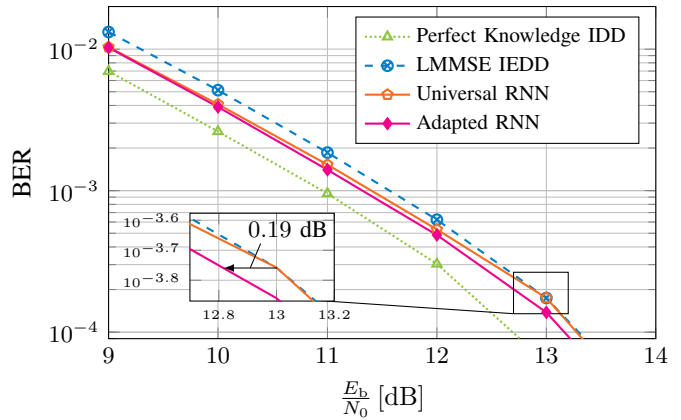}
    \end{center}
    \vspace{-2mm}
    \caption{\ac{BER} performance of the investigated receivers in the corner case scenario of no movement and thereby no channel time-variance ($v = \unitfrac[0]{km}{h}$ and moderate $L = 8$ channel taps).}
    \label{fig:BER_retraining_8taps}
    \vspace{-5mm}
\end{figure}

The first scenario investigates the impact of adaptation to corner case conditions using the example of no \ac{UE} movement.
For this purpose we set the velocity to $v = \unitfrac[0]{km}{h}$ and choose a moderate number of $L = 8$ channel taps so that the stochastic channel model generates channel realizations that form a sub-ensemble of the universal conditions used for initial training (Tab.~\ref{Tab:Training_Parameters}).
As can be seen from the results shown in Fig.~\ref{fig:BER_retraining_8taps}, the unadapted \emph{Universal RNN} already shows a better performance than the conventional \emph{LMMSE IEDD} baseline, thus, confirming the findings of \cite{aoudia2020end, Fischer_2021}.
This gain can be justified by the fact that the RNN-based receiver can additionally exploit the expected distribution of the data-carrying symbols in $\Ym$.
However, by adapting the RNN receiver to the current channel conditions, the \emph{Adapted RNN} can further gain about \unit[0.1]{dB} of BER performance compared to the \emph{Universal RNN}.
Interestingly, this gain is possible although the channel conditions of this scenario were part (sub-ensemble) of the initial universal training.
We assume that retraining to current channel conditions reinforces the RNN to lift conservative assumptions, as channel realizations with high velocity are not part of the retraining data and high velocity implications are thereby not considered for weight updates.
These gains have also been observed for various other corner cases with different parameters within the range of the universal channel ensemble, but due to paper length limits we exemplary only show this corner case.

\subsection{Out-of-Specification (Extreme) Scenario}

\begin{figure}[t]
    \begin{center}
 	  \input{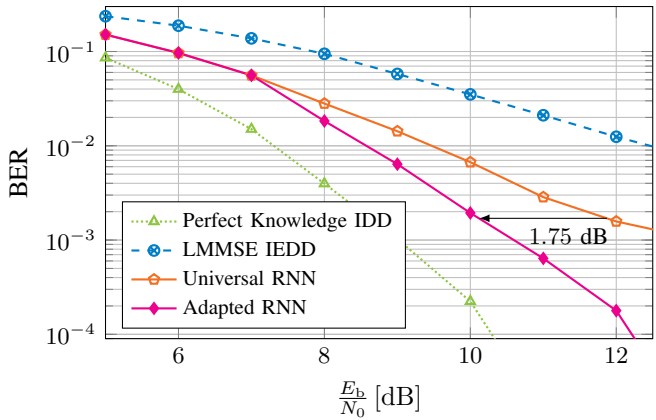}
    \end{center}
    \vspace{-2mm}
    \caption{\ac{BER} performance of the investigated receivers in the extremely frequency-variant (out-of-specifications) scenario of $L = 16$ channel taps at a moderate velocity of $v = \unitfrac[100]{km}{h}$.}
    \label{fig:BER_retraining_16taps}
    \vspace{-4mm}
\end{figure}

In the second scenario, we want to focus on the benefit of adaptation in case of unforeseen and extreme channel conditions.
Therefore, the results shown in Fig.~\ref{fig:BER_retraining_16taps} were obtained at highly frequency-selective channel conditions with $L = 16$ channel taps at a moderate velocity of $v=\unitfrac[100]{km}{h}$.
The simulation results show that the performance of the conventional \emph{LMMSE IEDD} baseline system degrades heavily.
This is expected as it mainly relies on pilot symbols and the used pilot position spacing in frequency dimension is not sufficient for $L = 16$ channel taps, setting this scenario out of specification.
Likewise, this scenario is also out of specification for the \emph{Universal RNN} as initial training only covers channel conditions up to $L = 14$ channel taps.
However, the performance of the \emph{Universal RNN} does also degrade compared to the \emph{Perfect Knowledge IDD} lower limit, but not as much as the \emph{LMMSE IEDD} baseline system.
This observation is also consistent with the findings of \cite{aoudia2020end, Fischer_2021} which showed, that NN-based receivers extract further knowledge about the channel from the provided data-carrying symbols and are therefore more robust against sparse pilot spacing.
But, most interestingly, the \emph{Adapted RNN} shows significantly improved performance compared to the \emph{Universal RNN}.
While there is still a large gap between the performance of the \emph{Adapted RNN} and \emph{Perfect Knowledge IDD}, these results show that adaptation can render a NN-based receiver to significantly higher operability, even in the case of a scenario that was originally out of specifications.

\subsection{Interference Scenario}

\begin{figure}[t]
    \begin{center}
 	    \input{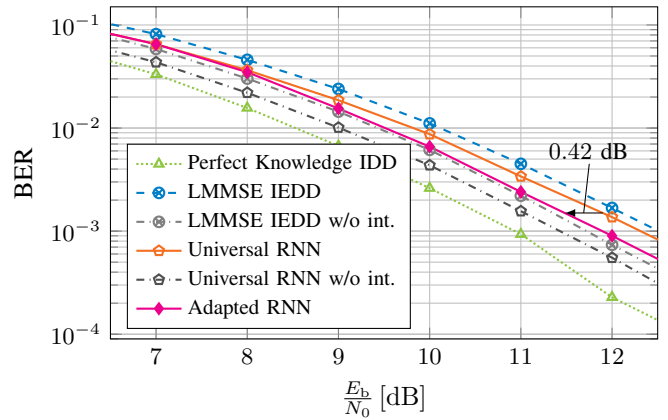}
    \end{center}
    \vspace{-2mm}
    \caption{
    \ac{BER} performance of the investigated receivers in a scenario with side channel interference, modeled by additive noise of $\unit[6]{dB}$ on the outer four subcarriers, at otherwise moderate conditions with $L = 8$ channel taps and $v = \unitfrac[100]{km}{h}$.}
    \label{fig:BER_retraining_guard_band_interference}
    \vspace{-4mm}
\end{figure}

Finally, we want to showcase a scenario that highlights the flexibility of NN-based receivers and how retraining can even enable adaptation to unseen tasks.
This is shown using the example of side channel interference, which is modeled by adding noise to the outer four subcarriers, reducing their SNR by $\unit[6]{dB}$.
As can be seen from the results shown in Fig.~\ref{fig:BER_retraining_guard_band_interference}, the \emph{LMMSE IEDD} baseline as well as the \emph{Universal RNN} suffer from the added interference, but retraining the RNN-based receiver leads to a performance gain of \unit[0.42]{dB} when we compare the \emph{Adapted RNN} with the \emph{Universal RNN}.
In this case the NN-based receiver is able to cope with the new task of incorporating the disturbance on the outer four subcarriers via retraining, while a conventional system would require additional signal processing and can not simply adapt.

\section{Conclusion}
\label{sec:conclusion}
We have demonstrated that \ac{NN}-based receivers benefit from continuous retraining as they can adapt to current, extreme and new unforeseen channel conditions.
For such cases, adaptation leads to a superior performance when compared to static receivers that have only been designed and optimized for a universal channel model.
Finally, we want to emphasize that these gains come without any additional signaling overhead, as on-the-fly label recovery is sufficient for the retraining process.

\bibliographystyle{IEEEtran}
\bibliography{IEEEabrv,references}

\end{document}